\documentclass[12pt]{article}

\begin{document}

\vspace{1pt} LOWEST OPEN CHANNELS, BOUND STATES, AND NARROW RESONANCES OF
DIPOSITRONIUM

\vspace{1pt}

C.G. Bao$^{1,2}$ and T.Y. Shi$^{2}$

\vspace{1pt}$^{1}$Center of Theoretical Nuclear Physics, National Laboratory
of Heavy Ion Accelerator, Lanzhou 730000, China

\vspace{1pt}$^{2}$Department of Physics, Zhongshan University, Guangzhou,
510275, China \vspace{1pt}

\vspace{1pt}

ABSTRACT: The constraints imposed by symmetry on the open channels of
dipositronium has been studied, and the symmetry-adapted lowest open channel
of each quantum state has been identified. Based on this study, the
existence of two more 0$^{+}$ bound states has been theoretically confirmed,
and a 0$^{+}$ narrow resonance has been predicted. \ A variational
calculation has been performed to evaluate the critical strength of the
repulsive interaction . Two 0$^{-}$ states are found to have their critical
strengths very close to 1, they are considered as candidates of new narrow
resonances or loosely bound states . \

\vspace{1pt}

1, INTRODUCTION

\qquad Since positrons can be easily created in the processes of high energy
collisions, the molecules formed by electrons and positrons are believed to
exist in nature. Half a century ago the existence of the positronium
molecule, namely the dipositronium Ps$_{2}$ . has already been predicted$^{1}
$ and has been firstly calculated via a variational procedure$^{2}$. Since
the lifetime of the ground state of Ps$_{2}$ is very short (it is only
0.906ns$^{3}$) due to the e$^{-}$-e$^{+}$ annihilation, it has not yet been
observed experimentally. In recent years \ this problem has attracted
increasing attention following the ability to create cold \ positron beams
tunable over a wide energy range$^{4}$, and the increasing ability to carry
out accurate theoretical calculations $^{3,5-10}$. In addition to the ground
state, an angular momentum L=1 and parity $\Pi =-1$ bound excited state has
also been predicted$^{10}.$ The other excited states are believed to be
resonances.

\qquad\ Of course any resonance would collapse via at least one open
channels. However, whether a state is allowed to enter into an open channel
is not only determined by energy but also by symmetry. \ In fact, as we
shall see, the\ states with a specific set of quantum numbers are allowed by
symmetry to get access to only a few specific open channels. Thus, for a
given state, it is important to make sure which open channel is the lowest
one allowed by symmetry, this is called a symmetry-adapted lowest open
channel (SLOC). If a state has its energy lower than the SLOC, it would be
definitely bound because it has no channel to collapse. If a state has its
energy a little higher than the SLOC and if there is a barrier to hinder the
wave function from leaking into the channel, it would appear as a narrow
resonance. The barrier may have different origin, what is concerned here is
the centrifugal barrier as discussed later. If the barrier does not exist or
if the energy is much higher than the SLOC, the state would be a broad
resonance. When a resonance has a very broad width, it is very difficult to
be detected, and thus the existence of this resonance is meaningless. In
other words, only the resonances with narrower widths are interesting.

\qquad The main aim of this paper is the identification of the SLOC, this is
the base for the further study of the dipositronium. Additionally, based on
the SLOC and on other existing data, the existence of a few more bound
states and resonances has been confirmed. Furthermore, a few candidates of
bound states or resonances has been suggested based on an evaluation. \ The
emphasis is placed on the qualitative aspect.

\vspace{1pt}

2, SYMMETRY-ADAPTED LOWEST OPEN CHANNELS

\qquad Let the particles 1 and 2 be electrons and 3 and 4 be positrons, let $%
{\bf r}_{a}{\bf =r}_{2}{\bf -r}_{1}$, ${\bf r}_{b}{\bf =r}_{4}{\bf -r}_{3}$,
${\bf r}_{c}{\bf =}\frac{1}{2}({\bf r}_{4}{\bf +r}_{3}$ ${\bf -r}_{2}{\bf -r}%
_{1})$ , and $r_{ji}{\bf =|r}_{j}{\bf -r}_{i}|$, The internal Hamiltonian of
the dipositronium is (in a.u.)

\vspace{1pt}

$H=-(\nabla _{a}^{2}+\nabla _{b}^{2}+\frac{1}{2}\nabla _{c}^{2})+(\frac{1}{%
r_{12}}+\frac{1}{r_{34}}-\frac{1}{r_{13}}-\frac{1}{r_{14}}-\frac{1}{r_{23}}-%
\frac{1}{r_{24}})\qquad (1)$

\vspace{1pt}

which is invariant with respect to the O$_{3}$ group and to the point group D%
$_{2d}$.$^{3}$ \ \ The latter is a subgroup of S$_{4}$ containing eight
elements, namely {\bf 1}, $p_{12}$, $p_{34}$, \ $p_{12}p_{34}$, \ $%
p_{13}p_{24}$, \ $p_{14}p_{23}$, $p_{c}$(1324), and $p_{c}$(1423) (where\ $%
p_{ij}$ denotes an interchange and $p_{c}$($ijkl$) denotes a cyclic
permutation). \ Hence the eigenstates can be classified according to the
irreducible representations of the O$_{3}$ and D$_{2d}$ , and thereby can be
denoted as $\ L^{\Pi }(\mu )$, where $\mu $ denotes a representation of the D%
$_{2d}$ \ group, $\mu =A_{1},$ $A_{2},$ $B_{1},$ $B_{2}$ or $E.$ Let the
wave function of a $L^{\Pi }(\mu )$ state be denoted as \ $\Psi _{L\Pi \mu }$%
. From the knowledge of the D$_{2d}$ group, we have$^{2}$

$p_{12}p_{34}$\ $\Psi _{L\Pi \mu }=$\ $\Psi _{L\Pi \mu }$ (if $\mu \neq E$),
or $=-$\ $\Psi _{L\Pi \mu }$ (if $\mu =E$)\qquad (2)

and

$p_{13}p_{24}$\ $\Psi _{L\Pi \mu }=$\ $\Psi _{L\Pi \mu }$ (if $\mu =A_{1}$
or $B_{1}$), or $=-$\ $\Psi _{L\Pi \mu }$ (if $\mu =A_{2}$ or $B_{2}$)
\qquad (3)

Let the spins of the two electrons 1 and 2 be coupled to $S_{1}$, those of 3
and 4 be coupled to $S_{2}$. Obviously, eq.(2) implies that the states with $%
\mu \neq E$ have $S_{1}=S_{2}$ .

\vspace{1pt}

\qquad On the other hand, let $\Phi _{nl}$ denotes an eigenstate of the
positronium Ps ( the ground state has $(nl)=(10)$ ). For the Ps$%
_{2}\rightarrow $Ps + Ps dissociation channel (where the Ps may be excited),
The channel wave function $\Psi _{ch}$\ can be written as

$\Psi _{ch}=(1+(-1)^{S_{1}}p_{12})(1+(-1)^{S_{2}}p_{34})\lbrack (\Phi _{nl}(%
{\bf r}_{13})\Phi _{n^{\prime }l^{\prime }}({\bf r}_{24}))_{l_{o}}f_{l_{R}}(%
{\bf r}_{13,24})\rbrack _{L}\qquad (4)$

where \ $f_{l_{R}}$\ \ is the wave function of \ relative motion and $l_{R}$
is the relative angular momentum. $l$ and $l^{\prime }$\ are coupled to $%
l_{o}$, together with $l_{R}$\ they are coupled to $L$. \ From (4) we have

$p_{13}p_{24}\Psi _{ch}=(-1)^{l+l^{\prime
}}(1+(-1)^{S_{2}}p_{12})(1+(-1)^{S_{1}}p_{34})\lbrack (\Phi _{nl}({\bf r}%
_{13})\Phi _{n^{\prime }l^{\prime }}({\bf r}_{24}))_{l_{o}}f_{l_{R}}({\bf r}%
_{13,24})\rbrack _{L}$

$\qquad (5)$

Furthermore, in the special case of $(nl)=(n^{\prime }l^{\prime })$ , eq.(4)
can be rewritten as

$\Psi _{ch}=(1+(-1)^{S_{1}+S_{2}+l_{o}+l_{R}})\{\lbrack (\Phi _{nl}({\bf r}%
_{13})\Phi _{nl}({\bf r}_{24}))_{l_{o}}f_{l_{R}}({\bf r}_{13,24})\rbrack
_{L} $

$\ \ \ \ \ \ \ \ \ \ \ \ \ \ \ \ \ \ \ \ \ \ \ \ \ +(-1)^{S_{1}}\lbrack
(\Phi _{nl}({\bf r}_{23})\Phi _{nl}({\bf r}_{14}))_{l_{o}}f_{l_{R}}({\bf r}%
_{23,14})\rbrack _{L}\}\qquad (6)$

\vspace{1pt}

Evidently, if a $\ L^{\Pi }(\mu )$ state is allowed to enter into a specific
Ps + Ps channel, the associated $\Psi _{ch}$\ must have the same ($L$, $\Pi $%
, $\mu $) symmetry.

\vspace{1pt}

\qquad As examples, let us first study the $0^{-}(A_{1})$ and \ $%
0^{-}(B_{1}) $ states. From eq.(3) we have

$p_{13}p_{24}\Psi _{ch}=\Psi _{ch}\qquad (7.1)$

Since the $\mu =A_{1}$ or $B_{1}$ states have\ $S_{1}=S_{2}$, eq.(5) can be
rewritten as

$p_{13}p_{24}\Psi _{ch}=(-1)^{l+l^{\prime }}\Psi _{ch}\qquad (7.2)$

\vspace{1pt}Comparing eq.(7.1) with (7.2), we have

$(-1)^{l+l^{\prime }}=1\qquad (8.1)$

Furthermore, since we have meanwhile\ $S_{1}=S_{2}$, the factor\ $%
(1+(-1)^{S_{1}+S_{2}+l_{o}+l_{R}})$ in eq.(6) is nonzero only if $%
l_{R}+l_{o} $\ is even. \ Thus we have

(-1)$^{l_{o}+l_{R}}$ =1, if $(nl)=(n^{\prime }l^{\prime })$ \qquad (8.2)

Besides, it can be proved that none of the partial waves of a 0$^{-}$ state
can be zero. The proof is as follows. Let the angular momenta corresponding
to the three Jacobian vectors {\bf r}$_{a}$, {\bf r}$_{b}$, and {\bf r}$_{c}$
be denoted as $l_{1}$, $l_{2}$, and $l_{3}$ . For 0$^{-}$ states, \ any
relative partial wave, say $l_{1},$ \ can not be zero. If $l_{1}$\ were
zero, \ $l_{2}$ must be equal to $l_{3}$ to assure L=0. However, if $l_{2}$=
$l_{3}$ , the parity must be even. Therefore $l_{1}$ (or any relative
angular momentum) is not allowed to be zero in 0$^{-}$ states. Evidently,
the proof \ is valid for any set of Jacobian coordinates. Thus we have

$l\neq 0,$ $l^{\prime }\neq 0,$ and $l_{R}\neq 0\qquad (8.3)$

Definitely, a Ps+Ps channel is accessible to the $0^{-}(A_{1})$ or \ $%
0^{-}(B_{1})$ states only if (8.1) to (8.3) are satisfied. Due to these
constraints, the lowest Ps + Ps channel is the one having $(nl)=(n^{\prime
}l^{\prime })=(21)$ , and $l_{R}=1$ . This channel is labeled as (21)$\frac{p%
} {}(21),$ where $p$ denotes the relative $p$-wave of the two excited
positronium, the associated threshold energy is -0.1250 (a.u. are used in
this paper).

\vspace{1pt}

\qquad For the $0^{-}(A_{2})$ and \ $0^{-}(B_{2})$ states, the constraint
eq.(8.1) should be changed to (-1)$^{l+l^{\prime }}$= -1 ( thus $l_{R}$ is
even), while the constraints (8.2) and (8.3) remain unchanged. Accordingly
the lowest Ps+Ps channel is the (21)$\frac{d} {}(32)$ channel at -0.090278.
\ From a similar deduction, we know \ that the lowest Ps+Ps channels for the
$0^{-}(E)$ states are the (21)$\frac{d} {}(32)$ channel and the (21)$\frac{p%
} {}(31)$ channel both at -0.090278.

\vspace{1pt}

\qquad Let us inspect another 2-body channel, the Ps$^{-}+e^{+}$ (or Ps$%
^{+}+e^{-}$ ) channel. It is recalled that the Ps$^{-}$ has only one bound
state with angular momentum zero. Thus, if a $0^{-}(\mu )$ state entered
into this channel, $l_{R}$ must be zero to assure L=0. However, $l_{R}=0$ is
prohibited due to eq.(8.3). Therefore the Ps$^{-}+e^{+}$ channel is not
accessible to the $0^{-}(\mu )$ states.

\vspace{1pt}

\qquad Let us inspect the Ps $+e^{+}+e^{-}$ 3-body channel. Let $l$ be the
angular momentum of the Ps. Since $l=0$ is not allowed due to eq.(8.3), The
positronium in the 3-body channel of 0$^{-}$ states must be excited,
therefore the lowest Ps $+e^{+}+e^{-}$ 3-body channel is at -0.0625 (in
which the Ps has $(nl)=(21)$).

\vspace{1pt}

\qquad The above analysis is straight forward to be generalized and thereby
all  the SLOC  can be identified as listed in Table 1 ( the states having
sufficient data from existing theoretical calculations are listed in this
table) and Table 2, ( otherwise).

Table 1, The SLOC and the corresponding threshold energies (in a.u.) of the $%
L^{\Pi }(\mu )$ states ($L\leq 2$) together with the eigenenergies from
theoretical calculations ( $a$\ is from ref.3 , $b$\ is from ref.10, and $c$%
\ is from our evaluation) . The partial wave given in the third column is
the lowest wave.

\vspace{1pt}
\begin{tabular}{|c|c|c|c|c|}
\hline
L$^{\Pi }$ & $\mu $ & SLOC & threshold energy & eigenenergy \\ \hline
0$^{+}$ & A$_{1}$ & (10)$\frac{s} {}$(10) & -0.50000 & -0.516003778$^{b}$, \
-0.509$^{c}$ \\ \hline
0$^{+}$ & B$_{1}$ & (10)$\frac{s} {}$(10) & -0.50000 & -0.4994428$^{a}$, \
-0.485$^{c}$ \\ \hline
0$^{+}$ & A$_{2}$ & (10)$\frac{p} {}$(21) & -0.31250 & -0.3120805$^{a}$,
-0.301$^{c}$ \\ \hline
0$^{+}$ & B$_{2}$ & (10)$\frac{p} {}$(21) & -0.31250 & -0.3144689$^{a}$,
-0.307$^{c}$ \\ \hline
0$^{+}$ & E & (10)$\frac{s} {}$(20), \ (10)$\frac{p} {}$(21) & -0.31250 &
-0.3300469$^{a}$, \ -0.322$^{c}$ \\ \hline
1$^{-}$ & B$_{2}$ & (10)$\frac{s} {}$(21) & -0.31250 & -0.334408$^{b}$, \ \
\ -0.326$^{c}$ \\ \hline
\end{tabular}

\vspace{1pt}

Table 2, The SLOC, the threshold energies, and the critical strength $%
\lambda _{o}$. The $\lambda _{o}$ that smaller than 0.9 is not listed.

\vspace{1pt}
\begin{tabular}{|c|c|c|c|c|}
\hline
L$^{\Pi }$ & $\mu $ & SLOC & threshold energy & $\lambda _{o}$ \\ \hline
0$^{-}$ & A$_{1}$ & (21)$\frac{p} {}$(21) & -0.12500 & 0.93 \\ \hline
0$^{-}$ & B$_{1}$ & (21)$\frac{p} {}$(21) & -0.12500 & 0.98 \\ \hline
0$^{-}$ & A$_{2}$ & (21)$\frac{d} {}$(32) & -0.090278 & 0.94 \\ \hline
0$^{-}$ & B$_{2}$ & (21)$\frac{d} {}$(32) & -0.090278 &  \\ \hline
0$^{-}$ & E & (21)$\frac{p} {}$(31), (21)$\frac{d} {}$(32) & -0.090278 & 0.99
\\ \hline
\end{tabular}

\begin{tabular}{|c|c|c|c|c|}
\hline
L$^{\Pi }$ & $\mu $ & SLOC & threshold energy & $\lambda _{o}$ \\ \hline
1$^{+}$ & A$_{1}$ & (10)$\frac{d} {}$(32) & -0.27778 &  \\ \hline
1$^{+}$ & B$_{1}$ & (10)$\frac{d} {}$(32) & -0.27778 &  \\ \hline
1$^{+}$ & A$_{2}$ & (10)$\frac{p} {}$(21) & -0.31250 &  \\ \hline
1$^{+}$ & B$_{2}$ & (10)$\frac{p} {}$(21) & -0.31250 &  \\ \hline
1$^{+}$ & E & (10)$\frac{p} {}$(21) & -0.31250 &  \\ \hline
1$^{-}$ & A$_{1}$ & (10)$\frac{p} {}$(20) & -0.31250 & 0.93 \\ \hline
1$^{-}$ & B$_{1}$ & (10)$\frac{p} {}$(20) & -0.31250 &  \\ \hline
1$^{-}$ & A$_{2}$ & (10)$\frac{s} {}$(21) & -0.31250 &  \\ \hline
1$^{-}$ & E & (10)$\frac{p} {}$(10) & -0.50000 & 0.90 \\ \hline
\end{tabular}

\begin{tabular}{|c|c|c|c|c|}
\hline
L$^{\Pi }$ & $\mu $ & SLOC & threshold energy & $\lambda _{o}$ \\ \hline
2$^{+}$ & A$_{1}$ & (10)$\frac{d} {}$(10) & -0.50000 &  \\ \hline
2$^{+}$ & B$_{1}$ & (10)$\frac{d} {}$(10) & -0.50000 &  \\ \hline
2$^{+}$ & A$_{2}$ & (10)$\frac{p} {}$(21) & -0.31250 &  \\ \hline
2$^{+}$ & B$_{2}$ & (10)$\frac{p} {}$(21) & -0.31250 & 0.94 \\ \hline
2$^{+}$ & E & (10)$\frac{d} {}$(20), \ (10)$\frac{p} {}$(21) & -0.31250 &
0.93 \\ \hline
2$^{-}$ & A$_{1}$ & (10)$\frac{p} {}$(32) & -0.27778 &  \\ \hline
2$^{-}$ & B$_{1}$ & (10)$\frac{p} {}$(32) & -0.27778 &  \\ \hline
2$^{-}$ & A$_{2}$ & (10)$\frac{d} {}$(21) & -0.31250 &  \\ \hline
2$^{-}$ & B$_{2}$ & (10)$\frac{d} {}$(21) & -0.31250 &  \\ \hline
2$^{-}$ & E & (10)$\frac{d} {}$(21) & -0.31250 &  \\ \hline
\end{tabular}

\vspace{1pt}

3, DISCUSSION OF THE SPECTRUM

\ \qquad (A) Confirmation of new bound states and resonance

\qquad Among the states of Table 1, the 0$^{+}$(A$_{1}$) and 1$^{-}$(B$_{2}$%
) have already been pointed out that they are bound.$^{3,10}$ For the 0$^{+}$%
(B$_{2}$) and 0$^{+}$(E) states, they are tentatively classified as resonant
singlet and triplet spin state in the first paper of ref. 7, and are
believed to be metastable or resonance states in ref.3. Now, it is clear
that these two states have their energies lower than their thresholds, thus
they are definitely bound.

\qquad For the 0$^{+}$(A$_{2}$) state, its energy is only a little higher
than the threshold. Furthermore, the lowest partial wave for the relative
motion of the two dissociating positroniums is $p-$wave as shown in Table 1.
Thus a \ centrifugal barrier exists to hinder the wave function from leaking
out. \ It is recalled that the height of the centrifugal barrier of the Ps +
Ps channel is $l_{R}(l_{R}+1)/(2r_{ch}^{2})$, where $r_{ch}$ is roughly the
sum of the radii of the two projectiles ( a positronium in the ground state
and\ the other one in the (21) excited state), and therefore would be in the
order of 10 . Accordingly the height of the barrier would be in the order of
0.01 for $p-$wave. \ On the other hand, the energy of the 0$^{+}$(A$_{2}$)
is only higher than the threshold by 0.0004.$^{3}$ Thus the barrier is
sufficiently high to hinder the wave function, and therefore we believe that
the width of the 0$^{+}$(A$_{2}$) resonance is very narrow. For the 0$^{+}$(B%
$_{1}$) state, its energy is also only a little higher than the threshold.
However, the lowest partial wave for the dissociation is $s-$wave, thus \
centrifugal barrier does not exist, and therefore the 0$^{+}$(B$_{1}$) is a
broad resonance difficult to detect.

\vspace{1pt}\qquad

(B) Candidates of new bound states and resonances

\qquad\ In order to evaluate the eigenenergies of the states in Table 2, a
set of basis functions is introduced \ for the diagonalization of the
Hamiltonian. It is well known that the harmonic oscillator (h.o.) states are
in general not appropriate  for Coulomb systems, the main shortcoming is
their inappropriate asymptotic behavior. However, if several set of h.o.
states with different widths are used together, this shortcoming can be
remarkably cured$^{11}$. For example, let \ $\varphi _{nl}^{\omega _{K}}$\
be the eigenstate of a single-particle pure harmonic oscillation with the
width $\omega _{K}$ , eigenenergy ($2n+l+3/2$)$\hbar \omega _{K}$ and
angular momentum $l\hbar $ .\ When five sets of \ $\varphi _{nl}^{\omega
_{K}}$\ are used together (i.e.,  $\omega _{K}$ has five choices), and $%
n\leq 3$ is assumed for diagonalizing the Hamiltonian of a hydrogen atom,
the resultant lowest energies for $l=0,1,$and $2$ states are -0.499987,
-0.124998, and -0.055554, respectively, to be compared with the exact values
-0.5, -0.125, and -0.055556. Therefore, we believed that, if only bound
states are taken into account, and if only the qualitative aspect is
concerned, several sets of h.o. states together can be used for our limited
purpose.

\vspace{1pt}

\ \qquad In the follows, the following seven sets of basis functions

\vspace{1pt}$\Phi _{K}^{\omega _{K}}=\lbrack \varphi _{n_{1}l_{1}}^{\omega
_{K}}({\bf r}_{a})\lbrack \varphi _{n_{2}l_{2}}^{\omega _{K}}({\bf r}%
_{b})\varphi _{n_{3}l_{3}}^{\omega _{K}}({\bf r}_{c})\rbrack _{l_{o}}\rbrack
_{L}\qquad $(9)

\vspace{1pt}

are used together, where $\omega _{K}=\omega _{1}$ to $\omega _{7}$ , $%
\omega _{i}/\omega _{i-1}=2.5$, and (-1)$^{l_{1}+l_{2}+l_{3}}$ = $\Pi $ . \
Based on \ $\Phi _{K}^{\omega _{K}}$\ , the basis functions of a specific
representation $\mu $ can be induced by the following idempotents of the D$%
_{2d}$ group$^{3}$.

$e^{A_{1}}=\frac{1}{8}$(1+$p_{13}p_{24}$)(1+$p_{12}$)(1+$p_{34}$) \ \ \ \ \
\ \ \ \ \ \ \ \ \ \ (10.1)

$e^{A_{2}}=\frac{1}{8}$(1-$p_{13}p_{24}$)(1 -$p_{12}$)(1 -$p_{34}$) \ \ \ \
\ \ \ \ \ \ \ \ \ \ \ (10.2)

$e^{B_{1}}=\frac{1}{8}$(1+$p_{13}p_{24}$)(1-$p_{12}$)(1 -$p_{34}$) \ \ \ \ \
\ \ \ \ \ \ \ \ \ \ (10.3)

$e^{B_{2}}=\frac{1}{8}$(1-$p_{13}p_{24}$)(1+$p_{12}$)(1+$p_{34}$) \ \ \ \ \
\ \ \ \ \ \ \ \ \ \ (10.4)

$e^{E_{11}}=\frac{1}{4}$(1-$p_{12}p_{34}$+$p_{13}p_{24}$ -$p_{14}p_{23}$) \
\ \ \ \ \ \ \ \ \ \ \ \ \ \ (10.5)

$e^{E_{22}}=\frac{1}{4}$(1-$p_{12}p_{34}$ -$p_{13}p_{24}$+$p_{14}p_{23}$) \
\ \ \ \ \ \ \ \ \ \ \ \ \ \ (10.6)

\vspace{1pt}Besides, a subsidiary procedure is needed to extract a new set \{%
$\Psi _{i}$\} from the old sets, so that the $\Psi _{i}$ are
orthonormalized. The total number of \ $\Psi _{i}$ used for the
diagonalization is determined by N$_{\max }$ which is the maximum of the sum
$2(n_{1}+n_{2}+n_{3})+l_{1}+l_{2}+l_{3}$ . As examples, when N$_{\max }$=23,
the number of $\Psi _{i}$\ for the 0$^{-}$($A_{1}$) and 0$^{-}$($B_{1}$)
states are 6655 and \ 7824 respectively.

\qquad \vspace{1pt}

\qquad Evidently, the above basis functions are designed for bound states
but not for resonances.\ \ \ However, most states in Table 2 are resonances
(as we shall see). \ In order to avoid such an awkward situation, we
introduce an adjustable parameter $\lambda $ in\ the repulsive interactions
as $\lambda (\frac{1}{r_{12}}+\frac{1}{r_{34}})$ ,while the attractive
interactions remain unchanged. Such an adjustment does not change the
threshold energies of the Ps+Ps channels. It is emphasized that the
introduction of $\lambda $ does not at all alter the symmetry of the
Hamiltonian, it remains to be invariant with respect to O$_{3}$ and D$_{2d}$
. Therefore the nature of the problem is not altered. Of course, what we
really concern is the case with $\lambda $ close to 1.

\qquad Let the lowest eigenenergy of a given ($L\Pi \mu $) symmetry be
denoted as $E(\lambda )$. Then, our procedure is firstly to choose a $%
\lambda $ smaller than one \ so that $E(\lambda )$ is lower than the
threshold of the SLOC. For an example, for the 0$^{-}$($B_{1}$) state, when $%
\lambda =0.95$, we have $E(\lambda )=-0.1282$ which is lower than the SLOC
at -0.1250. \ Secondly, \ $\lambda $\ is increased step by step (in each
step $\lambda $ is increased by 0.01). When $\lambda $ is equal to a value $%
\lambda _{o}$\ so that $E(\lambda _{o})$ is lower than while $E(\lambda
_{o}+0.01)$ is higher than the threshold, then the procedure stops and $%
\lambda _{o}$ is call a critical strength. \ In this way mainly bound states
are concerned in the calculation.\ Evidently, if a state has a $\lambda _{o}$
much larger than one, it is deeply bound. If $\lambda _{o}$ is a little
larger than one, it is just bound and a little lower than the dissociation
threshold. If $\lambda _{o}$ is a little smaller than one, it is a resonance
a little higher than the threshold. If $\lambda _{o}$ is remarkably smaller
than one, the state is a high-lying resonance with a broad width, and we
will neglect it.

\qquad To show the convergency of our calculation, the eigenenergies of
selected states are listed at the last column of Table 1 (with $\lambda =1$%
). Besides, two series of eigenenergies are given in Table 3 in accord with N%
$_{\max }$ . Owing to the difficulty in calculation, the basis
functions with N$_{\max }${>}23 are not adopted. The above results
exhibit that the speed of convergency is not good, \ thus the
calculated energies would not be useful if we want to know the
exact locations of the levels (this is an effort beyond the aim
and  scope of this paper). However, if we just want to \ know
which levels are relatively closer to their SLOC, then our results
are useful.

\vspace{1pt}

Table 3, The calculated energies $E(\lambda )$ (in a.u.) of the lowest $%
0^{-}(A_{1})$ state with $\lambda =1$ and the lowest $0^{-}(B_{1})$ state
with $\lambda =0.95$ when N$_{\max }$ is given.

\begin{tabular}{|l|l|l|l|l|l|}
\hline
N$_{\max }$ & 15 & 17 & 19 & 21 & 23 \\ \hline
$0^{-}(A_{1})$ & -0.1088 & -0.1110 & -0.1128 & -0.1142 & -0.1153 \\ \hline
$0^{-}(B_{1})$ & -0.1258 & -0.1267 & -0.1274 & -0.1279 & -0.1283 \\ \hline
\end{tabular}

\vspace{1pt}

\qquad The numerical results of $\lambda _{o}$\ are summarized in Table 2. \
In this table most $\lambda _{o}$ are remarkably smaller than one, the
associated states are broad resonances and will not be further discussed.
However, there are two and only two states, namely the $0^{-}(E)$ and \ $%
0^{-}(B_{1})$ , have their $\lambda _{o}$ very close to one. They are
distinguished from the others. Since, as mentioned before, all the
calculated energies are a little higher than the actual values, the actual
critical strength $\lambda _{actual}$ will be a little larger than the $%
\lambda _{o}$ listed in the table. Thus, the $\lambda _{actual}$ of the $%
0^{-}(E)$ and \ $0^{-}(B_{1})$ will be  in fact very close to one or even
larger. Furthermore, the open channels of these two states have $l_{R}$ to
be $p-$wave or higher, thus a centrifugal barrier exists. Therefore, among
all the other states, these two states are candidates, they are either
narrow resonances or bound states. If they are resonances, the $0^{-}(E)$
would emerge when two excited positroniums collide with each other (via the
(21)$\frac{p} {}$(31)\ \ or (21)$\frac{d} {}$($32$)\ channel). Similarly,
the $0^{-}(B_{1})$ would emerge in the collision via the (21)$\frac{p} {}$%
(21) channel. Once they are created, they are free from direct annihilation.$%
^{12}$ They would collapse either via their channel of formation, or would
be transformed to a lower broad resonance via an E1 transition (the $0^{-}(E)
$ would then become a $1^{+}(E)$, and the $0^{-}(B_{1})$ become a \ $%
1^{+}(A_{2})$), and then collapse via the (10)$\frac{p} {}$(21)\ channel.
Anyway, during the decay of these resonances, low-energy photons (from the
above E1 transition and from the transition of an excited positronium to its
ground state) can be detected. These low-energy photons would help the
identification of these states.

\vspace{1pt}

4, FINAL REMARKS

\qquad In conclusion, the main outcome of this paper is the
identification of the SLOC of dipositronium. The identification is
proved to be very useful to the analysis of the existing data.
Specifically, in addition to the two bound states that have been
found previously, two more 0$^{+}$ bound states have been
identified, and a 0$^{+}$ narrow resonances has been predicted.
Furthermore, two 0$^{-}$ states are found to have their energies
very close to their SLOC (while the others are remarkably higher).
Besides, their SLOC\ contain centrifugal barrier. Thus they are
either narrow resonances or loosely bound states. Since our
calculation is not accurate enough, what they really are remains
to be identified . Undoubtedly, the final identification of these
two states is an attractive theoretical topic. \ In the
experimental aspect, the search of the $0^{+}(A_{2})$ resonance
via the collision taking place in the (10)$\frac{p} {}$(21)
channel is firstly recommended, because its width is very narrow.

\qquad\ As a summary, a primary spectrum of dipositronium is proposed as
shown in Table 4, where at least four bound states have been identified.

\vspace{1pt}

Table 4, Bound states (b) and narrow resonances (r) of
dipositronium together with their energies (in a.u.). The state
lying upper is higher in energy. $\triangle $ is a small quantity
(positive or negative) in the order of 0.001. The values of the
energies of the 0$^{+}$ and 1$^{-}$ states come from the
literatures cited in Table 1.

\begin{tabular}{|l|l|l|}
\hline $L^{\Pi }(\mu )$ & energy &  \\ \hline $0^{-}(E)$ &
-0,090278+$\triangle $ & r or b \\ \hline $0^{-}(B_{1})$ &
-0.1250+$\triangle $ & r or b \\ \hline $0^{+}(A_{2})$ &
-0.3120805 & r \\ \hline $0^{+}(B_{2})$ & -0.3144689 & b \\ \hline
$0^{+}(E)$ & -0.3300469 & b \\ \hline $1^{-}(B_{2})$ & -0.334408 &
b \\ \hline $0^{+}(A_{1})$ & -0.516003778 & b \\ \hline
\end{tabular}

\vspace{1pt}

\qquad The above procedure is mainly based on symmetry consideration, the
way of analysis can be generalized to study other molecules with symmetries
other than O$_{3}$ and D$_{2d}$. In particular, the identification of the
SLOC is important to the study of resonances of various molecules.

\vspace{1pt}

A{\small cknowledgment: This work is supported by the NSFC of China under
the grants No.90103028 and No.10174098.}

\vspace{1pt}\vspace{1pt}

REFERENCES

1, A. Wheeler, Ann. N.Y. Acad. Sci. 48, 219 (1946)

2, E.A. Hylleraas and A. Ore, Phys. Rev. 71, 493 (1947)

3, D.B. Kinghorn and R.D. Poshusta, Phys. Rev. A, 47, 3671 (1993)

4, S.J. Gilbert, C. Kurz, R.G. Greves, and C.M. Surko, \ Appl. Phys. Lett.
70, 1944 (1997)

5, P.M. Kozlowski and L.Adamowicz, Phys. Rev. A 48, 1903 (1993).

6, A.M. Frolov, S.I. Kryuchkov, and V.H. Smith, Jr., Phy. Rev. A 51, 4514
(1995)

7, Y.K. Ho, Phys. Rev. A 39, 2709 (1989); A 48, 4780 (1993)

8, J.-M. Richard, Phys. Rev. A 49, 3573 (1994)

9, D. Bressanini, M. Mella, and G. Morosi , Pyhs. Rev. A 55, 200 (1997)

10, K. Varga, J. Usukura, and Y. Suzuki, Phys. Rev. Lett. 80, 1876 (1998)

11, M. Kamimura and H. Kameyama, Nucl. Phys. A508, 17c (1990)

12, When only the two-photon process is assumed, the probability of
annihilation is proportional to the probability of overlap of an electron
with a positron $\langle \delta ({\bf r}_{e^{+}}{\bf -r}_{e^{-}})\rangle $.
If this quantity is zero, the associated state is free from direct
annihilation. On the other hand, it has been stated in the text that
s-partial wave is not allowed in 0$^{-}$ states. However, wave function of
any partial wave must be zero at \ r=0, except the s-wave. Thus, the
prohibition of the s-partial \ wave leads to $\langle \delta ({\bf r}_{e^{+}}%
{\bf -r}_{e^{-}})\rangle $=0 and therefore the 0$^{-}$ states are free from
direct annihilation.

\vspace{1pt}

\end{document}